\documentclass[runningheads]{llncs}
\usepackage{graphicx}
\usepackage{url}
\usepackage{amsmath}
\usepackage{multirow}
\usepackage{diagbox}
\usepackage{color}
\begin{document}
\title{MS-NAS: Multi-Scale Neural Architecture Search for Medical Image Segmentation}
\titlerunning{MS-NAS: Multi-Scale Neural Architecture Search}
\author
{Xingang Yan\inst{1} \and
Weiwen Jiang\inst{2} \and
Yiyu Shi\inst{2} \and
Cheng Zhuo\inst{1}}

% 1 {Yan,Xingang}
% 2 {Jiang,Weiwen}
% 3 {Shi,Yiyu}
% 4 {Zhuo,Cheng}

\authorrunning{X. Yan et al.}

\institute{ZheJiang University, Hangzhou, China \\
\email{\{21832147,czhuo\}@zju.edu.cn}\\
\and
University of Notre Dame, Notre Dame, USA\\
\email{\{wjiang2,yshi4\}@nd.edu}}
\maketitle              % typeset the header of the contribution
\begin{abstract}
The recent breakthroughs of {Neural Architecture Search} (NAS) have motivated various applications in medical image segmentation. However, most existing work either simply rely on hyper-parameter tuning or stick to a fixed network backbone, thereby limiting the underlying search space to identify more efficient architecture. This paper presents a \emph{Multi-Scale NAS} (MS-NAS) framework that is featured with multi-scale search space from network backbone to cell operation, and multi-scale fusion capability to fuse features with different sizes. To mitigate the computational overhead due to the larger search space, a partial channel connection scheme and a two-step decoding method are utilized to reduce computational overhead while maintaining optimization quality. Experimental results show that on various datasets for segmentation, MS-NAS outperforms the state-of-the-art methods and achieves 0.6-5.4\% mIOU and 0.4-3.5\% DSC improvements, while the computational resource consumption is reduced by 18.0-24.9\%.

\end{abstract}

\section{Introduction}
Accurate segmentation of medical images is a crucial step in computer-aided diagnosis, surgical planning and navigation~\cite{ref_article_1}. The recent breakthroughs in deep learning, such as UNet~\cite{ref_article9}, have steadily improved segmentation efficiency, which not only defeats human visual systems, but also exceeds the conventional algorithms in both speed and accuracy~\cite{ref_article0,ref_article9,ref_article10}. However, in general, designers have to spend significant efforts through manual trial-and-error deciding network architecture, hyper-parameters, pre- and post-processing procedures~\cite{ref_article12}. Thus, it is highly desired to have an efficient network design procedure when segmenting for different modalities, subjects, and resolutions~\cite{ref_article11}.

The recently proposed automated machine learning (AutoML) is well aligned with such demands to $automatically$ design the neural network architecture instead of relying on human experiences and repeated manual tuning. More importantly, many works in \emph{Neural Architecture Search} (NAS) have already identified more efficient neural network architectures in general computer vision tasks~\cite{ref_article4,ref_article12}. Such success has motivated various NAS applications in medical image segmentation~\cite{ref_article5,ref_article6,ref_article7,ref_article11}. However, most of them either simply apply Darts-like framework \cite{ref_article2} with hyper-parameter tuning, or sticks to a fixed network backbone (e.g., UNet), thereby restricting the underlying optimization space to identify more efficient architectures for different modalities, such as CT, MRI, and PET~\cite{ref_article5,ref_article6,ref_article7}.

In addition, as medical images are featured with inhomogeneous intensity, similar floorplan, and low semantic information, it is then a natural idea to utilize both high-level semantics and low-level features as a combined effort (i.e., multi-scale fusion) to suffice segmentation efficiency. The effectiveness of multi-scale fusion has already been partially proved by UNet~\cite{ref_article9}, which fuses the features of the same sizes from encoders/decoders. Thus, most UNet-based NAS work implicitly embeds such fusion capability. However, the implicit embedding also restricts the fusion only to the features of the same size, thereby functioning as enforced operation and preventing further optimization. Intuitively speaking, $real$ multi-scale fusion should fuse features of different sizes to provide more informative content for segmentation efficiency. To address the aforementioned concerns, this paper proposes a \textbf{Multi-Scale NAS} (MS-NAS) framework to design neural network for medical image segmentation, which is featured with:
\begin{itemize}
\item \textbf{Multi-scale search space:} The framework employs a $larger$ search space at different scales, from network backbone, artificial module and cell, to operation, which can identify more optimal architecture for different tasks.
\item \textbf{Multi-scale fusion:} The framework also explores the $real$ multi-scale fusion operation to improve segmentation efficiency by concatenating features at different scales within each artificial module.  
\end{itemize}
The proposed MS-NAS framework is an end-to-end solution to automatically determine the network backbone, cell type, operation parameters, and fusion scales. To facilitate such search, three types of cells, {\textit{expanding cells}}, {\textit{contracting cells}}, and {\textit{non-scaling cells}}, are defined in the next section to compose the learned network architecture. With the optimized cell types, fusion scales, and operation connections, the framework can identify among various backbones, such as UNet, ResUNet, FCN, etc., the most effective architecture meeting the varying demands from modality to modality. Thus, our proposal is different to the prior NAS work for medical image segmentation, which often sticks to one network backbone and only fuses the features of the same scale~\cite{ref_article5,ref_article6,ref_article7}. Apparently, the proposed MS-NAS can be resource consuming when optimizing in such a large search space. We here employ a partial channel connections scheme~\cite{ref_article8} and a two-step decoding method to speed up the search procedure. As a result, the proposed MS-NAS framework is capable to optimize on various high-resolution segmentation datasets in larger search space with reduced computational cost. 

%Then in the optimization and decoding stage, similar continuous relaxation and softmax sampling techniques as in \cite{ref_article4} are used to implement the discrete network architecture. 
Experimental results show that the proposed MS-NAS outperforms the prior NAS work~\cite{ref_article4,ref_article7} with 0.6-5.4\% mIOU and 0.4-3.5\% DSC improvement on average for various datasets while it achieves 18.0-24.9\% computational cost reduction. It is noted that the framework can trade off between accuracy and complexity, thereby enabling desired flexibility to build networks for different tasks while maintaining the state-of-the-art performance.

\begin{figure}[th]
\centering\vspace{-0.3cm}
\includegraphics[width=0.9\textwidth, height=0.65\textheight]{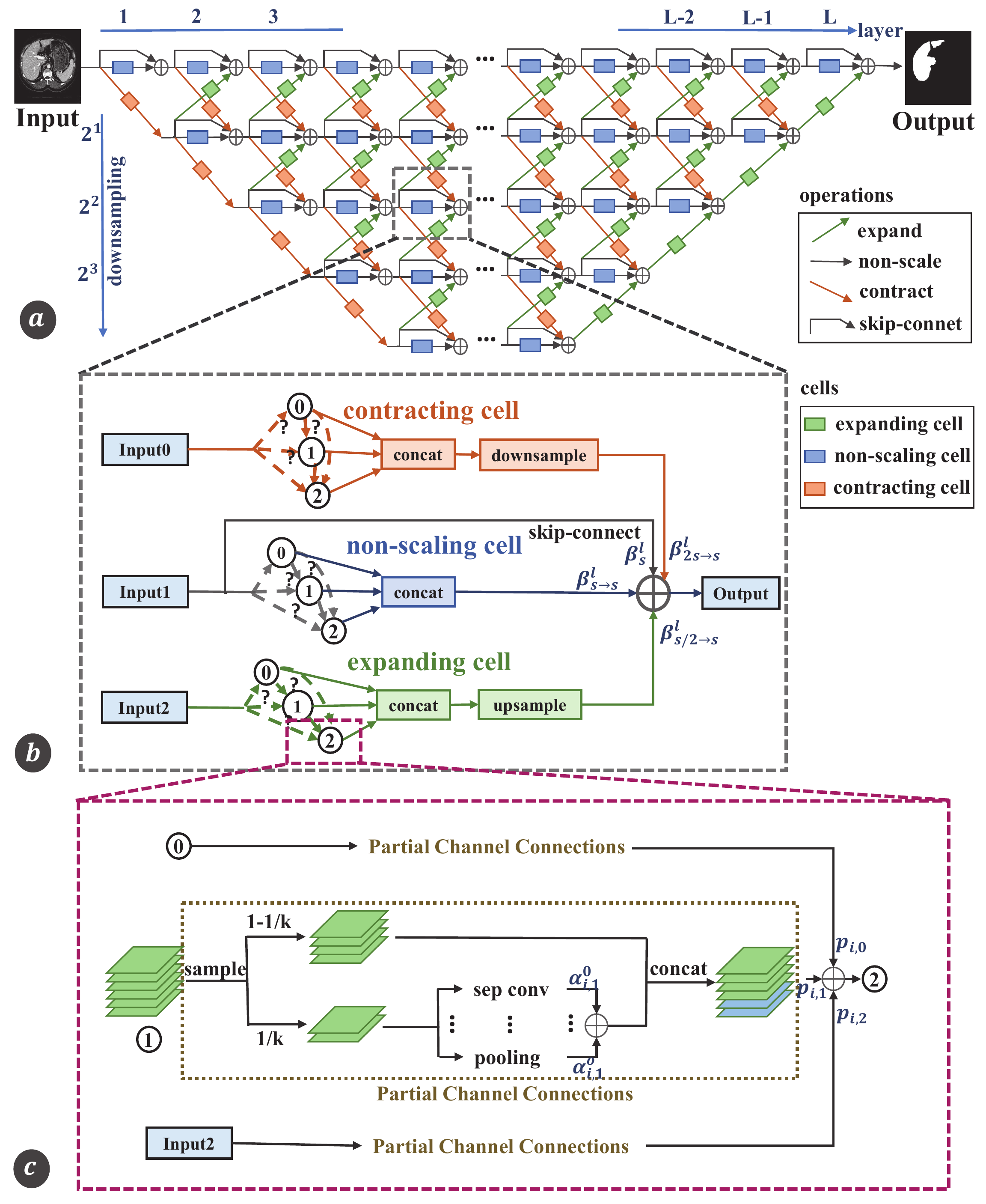}\vspace{-0.5cm}
\caption{Overview of the proposed architecture search space for medical image segmentation: (a) Search space for MS-NAS; (b) One artificial module containing three cells; and (c) Example illustration of partial channel connections.}\vspace{-0.5cm}
\label{fig:search_network}
\end{figure}
\section{Method}
\label{sec:method}
%The proposed MS-NAS framework contains two key steps, multi-scale search space structuring and contiuous relaxation, and optimization and decoding to the discrete architecture. 
In this section, we first present the proposed multi-scale architecture search space for medical image segmentation. Then, we discuss optimization and decoding schemes to obtain the discrete architecture. 
Fig.~\ref{fig:search_network}(a) provides an overview of the search space for MS-NAS as well as its components, which can be represented by a directed acyclic graph (DAG), with $V$ vertice and $E$ edges. Before we go into algorithm details, we would like to define the notations and components of the network from top to bottom for better discussions in the following sections. 

For the searched network, a \textit{sub-network} is defined as a path from input to segmentation output. Within the sub-network, the basic component is \textit{artificial module} (as shown in Fig.~\ref{fig:search_network}(b)), which may contain three types of $cells$: (1) \textit{Expanding cell} expands and up-samples the scale of feature map; (2) \textit{Contracting cell} contracts and down-samples the scale of feature map; and (3) \textit{Non-scaling cell} keeps the scale of feature map constant. With $module$ and $cell$ defined, we then define $operations$ that connect different cells within the modules and from module to module. In addition to the commonly used operations, such as $pooling$, we can add three $operations$, expanding, contracting, and non-scaling, corresponding to the three cells above. A \textit{skip-connect} operation within the cell is also employed to transfer the shallow features to the deep semantics. \vspace{-0.1cm}

\subsection{Multi-Scale Architecture Search Space}
\begin{figure}[htbp]
\centering
\includegraphics[width=\textwidth]{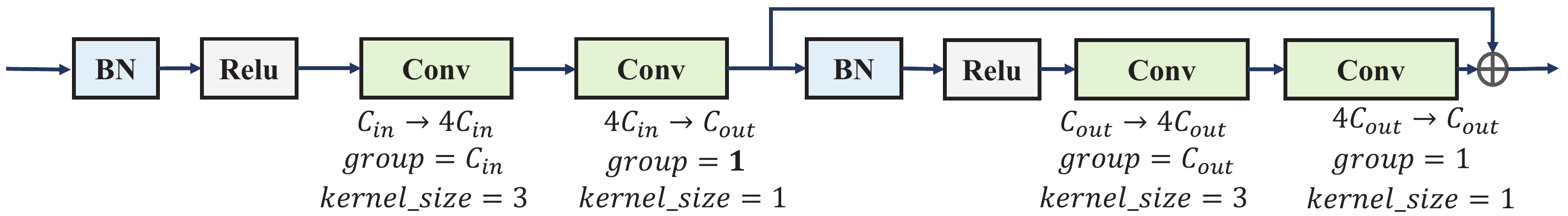}\vspace{-0.3cm}
\caption{An operator example of 3 $\times$ 3 \textit{depth-wise-separable convolution}, where $C_{in}$ and $C_{out}$ denote the number of channels for input and output tensors of a block.}\vspace{-0.5cm}
\label{fig:operation}
\end{figure}

The search space of MS-NAS covers different scales, from network, module, cell, to operation. \textit{Cell level} search is conducted at $local$ to find the desired cell structures and connections from the cell search space. Each cell search space can be represented by a DAG consisting of $N$ blocks, with each block representing the mapping from an input tensor $X_{in}$ to an output tensor $X_{out}$. For the $i_{th}$ block in a cell, we define a tuple ($I_i$,$o_i$) for such mapping: $I_i \in \mathcal{I}$, where $\mathcal{I}$ denotes a set of $X_{in}$ and output tensors {$X_1,…X_{i-1}$} from the $1_{st}$ to the $i-1_{th}$ blocks; and $o_i\in\mathcal{O}$ is the operation applied to $I_i$, where $\mathcal{O}$ denotes a set of operators modified to facilitate search, including {depth-wise-separable convolution}, {dilated convolution with a rate of 2}, {average pooling}, {skip connection}, etc. An operator example of 3$\times$3 \textit{depth-wise-separable convolution} is shown in Fig.~\ref{fig:operation}, with slightly changed operation order and additional ResNet-based skip connection. To reduce the memory cost during searching, a \textit{partial channel connections} scheme~\cite{ref_article8} is embedded in the framework. In particular, the channels of the input tensor for a block are partitioned to two parts according to a hyper-parameter $k$. The output tensor of the $i_{th}$ block then can be calculated as:\vspace{-0.15cm}
\begin{equation}X_{i}=\sum_{X_{j} \in \mathcal{I}_{i}} \frac{\exp \left\{p_{i, j}\right\}}{\sum_{j^{\prime} < i} \exp \left\{p_{i, j^{\prime}}\right\}} \cdot f_{i, j}^{\mathrm{PC}}\left(\mathrm{X}_{j} ; \mathrm{K}_{i, j}\right)\vspace{-0.1cm} \end{equation}
where $(i,j)$ denotes the edge connecting blocks $i$ and $j$, which is parameterized by a scalar $p_{i,j}$. $f_{i, j}^{\mathrm{PC}}\left(\mathrm{X}_{j} ; \mathrm{K}_{i, j}\right)$ is an auxiliary function for partial channel connection:\vspace{-0.1cm}
\begin{equation}f_{i, j}^{\mathrm{PC}}\left(\mathrm{X}_{j} ; \mathrm{K}_{i, j}\right)=\sum_{o \in \mathcal{O}} \frac{\exp \left\{\alpha_{i, j}^{o}\right\}}{\sum_{o^{\prime} \in \mathcal{O}} \exp \left\{\alpha_{i, j}^{o^{\prime}}\right\}} \cdot o\left(\mathrm{K}_{i, j} * \mathrm{X}_{j}\right)+\left(1-\mathrm{K}_{i, j}\right) * \mathrm{X}_{j}\vspace{-0.1cm}\end{equation}
where ${K}_{i, j}$ is a channel sampling mask. As shown in Fig.~\ref{fig:search_network}(c), $1/k$ portion of channels go through the operations selected from $\mathcal{O}$ while the rest remain unchanged. $\alpha_{i, j}^{o}$ parameterizes the operator in partial channel connection to control the contributions from different operators. This scheme helps reduce memory consumption during search while still maintaining a good convergence rate~\cite{ref_article8}. Finally, to weight the contributions from different edges when computing $X_i$, we use $\frac{\exp \left\{p_{i, j}\right\}}{\sum_{j^{\prime} < i} \exp \left\{p_{i, j^{\prime}}\right\}}$ for edge normalization. The output $X_{out}$ of the cell is then computed as the concatenation of the tensors {$X_1,X_2,…X_N$} for the $N$ blocks.

\textit{Network level} search is conducted to find the desired network backbone within the entire network search space to determine the network connection of sub-networks. As shown in Figure~\ref{fig:search_network}(a), the search space structure is gradually shrunk from top to bottom. Each path from input to output is unique and goes through different modules and different operations, resulting in different feature map scale changes. The search is then to find one or multiple sub-networks as well as their connections for the given hyper-parameters. 

To facilitate the search procedure, we use continuous relaxation for network search~\cite{ref_article2}. With $Ep(*)$, $Ns(*)$, $Ct(*)$ defined as the operators for expanding, non-scaling and contracting, the connection between the cells at layer $l$ can be parameterized by a scalar $\beta_{s_{1}\rightarrow {s_2}}^{l}$ for the three operators, where $s_1$, $s_2$ indicate the sampling scale\footnote{Without loss of generality, we use a factor of 2 for up- and down-sampling.}. A \textit{skip-connect} operations is parameterized by a scalar $\beta_{s}^{l}$ without an explicit operator. Then the output feature map for the $l_{th}$ layer is:\vspace{-0.1cm}
\begin{equation}
\begin{aligned}
X^{l}_{s}=
&Softmax(\beta_{\frac{s}{2} \rightarrow s}^{l})Ep(X^{l-1}_{\frac{s}{2}};\alpha_{ep},p_{ep})\\
&+ Softmax(\beta_{s \rightarrow s}^{l})Ns(X^{l-1}_{s};\alpha_{ns},p_{ns})\\
&+ Softmax(\beta_{2s \rightarrow s}^{l})Ct(X^{l-1}_{2s};\alpha_{ct},p_{ct})\\
&+ Softmax(\beta_{s}^{l})X^{l-1}_{s}
\end{aligned}\label{eq_edge}\vspace{-0.1cm}
\end{equation}
where $s \in \{2^0,2^1,2^2,2^3,2^4\}$ is the sacling ratio, and $ Softmax(\beta) = \frac{\exp \left\{\beta \right\}}{\sum_{\beta \in B} \exp \left\{\beta\right\}} $. Note that zero operation is also accounted in our framework, which is simply disconnection between the blocks.

\subsection{Optimization and Decoding}
With the defined search space and relaxed parameters, we formulate the architecture search as an continuous optimization problem, similar to \cite{ref_article2,ref_article4}. This can be effectively solved using stochastic gradient decent (SGD) method to obtain an approximate solution by optimizing the parameter sets of $\alpha$, $\beta$ and $p$ as discussed in the previous subsection~\cite{ref_article4}. Note that even with a larger search space than prior NAS work, the embedding of partial channel connection helps significantly reduce memory usage and time to make the framework feasible for desired high-resolution image segmentation tasks. This will be demonstrated in the experimental results in Section \ref{Experiments}.

After architecture search, we still need to derive the final architecture from the relaxed variables. Due to the unique multi-scale nature of our framework, we here propose a two-step decoding approach for cell and network structure determination. In the first step, at the cell level, the normalized coefficients $\frac{\exp \left\{p_{i, j}\right\}}{\sum_{j^{\prime} < i} \exp \left\{p_{i, j^{\prime}}\right\}}$ and $\frac{\exp \left\{\alpha_{i, j}^{o}\right\}}{\sum_{o^{\prime} \in \mathcal{O}} \exp \left\{\alpha_{i, j}^{o^{\prime}}\right\}}$ are multiplied as the weight for each edge $(i,j)$. Then the cell structure is identified by taking the operation associated to the edge with the highest weight. Simply extending this strategy to network structure is sub-optimal as the network has much larger size than a cell. On the other hand, conventional discrete optimization to decode the discrete architecture, such as dynamic programming, is infeasible due to a too high complexity of at least $O(V^2)$. Thus, here we propose a method to convert the network structure decoding to the top $N_l$ longest path search, where $N_l$ is the number of paths. 

In the second step, at the network level, the selection of top $N_l$ longest paths is based on the accumulated weights of all the edges in one path from input to output.
As in Eq.~\eqref{eq_edge}, we use parameters $\beta$ and $softmax$ function to formulate edge weights for the network. Kindly note that after the optimization, the sum of weights on the edges entering into a cell is always 1, which reflects the probability of strength or optimality. 
Inspired from this, the optimality or performance of a sub-network (corresponding to one path from input to output) can be partially measured by the sum of edge weights along the path.
Therefore, to identify $N_l$ top sub-networks is equivalent to find $N_l$ longest paths in the DAG graph.
This can be effectively solved by \textit{Dijkstra} algorithm with a complexity of $O(E\dot\log V)$.
Kindly note that the larger $N_l$ indicates more sub-networks to be included in the architecture, which is helpful in improving segmentation performance. As far as we know, this is the first work to incorporate MULTIPLE paths founded in a SuperNet (containing all searched paths) to make a better tradeoff between accuracy and hardware efficiency. Thus, the accuracy can approach that of SuperNet, while achieving high hardware efficiency. 
Experimental results in Section \ref{Experiments} will show that a small $N_l$ can achieve higher performance than UNet, with less Flops.

% We introduce $N_l$ for designers to give them freedom to deal with the trade offs between segmentation performance and hardware usage.

The proposed two-step decoding method not only has the ability to identify a high-quality network architecture, but also provides freedom for designers to make trade-offs between segmentation performance and hardware usage by adjusting hyper-parameter $N_l$.

\section{Experiments}\label{Experiments}

\subsection{Dataset and Experiment Setup}
We employ three datasets from Grand Challenges to evaluate the proposed MC-NAS, including: (1) Sliver07~\cite{ref_article14} dataset (liver CT scans, 8318 images, 20 training cases); (2) Promise12~\cite{ref_article15} dataset (prostate MRI scans, 1377 images, 20 training cases); and (3) Chaos~\cite{ref_article16} dataset (liver, kidneys and spleen MRI scans, 1270 images, 20 training cases). The Chaos dataset also contains CT scans for livers, which is used for MS-NAS to search a network architecture. The searched architecture is then adapted to each datasets with limited training cases as reported through transfer learning. The performance of the proposed MS-NAS is compared with several state-of-the-art NAS frameworks, including AutoDeeplab, which is considered as one of the best NAS frameworks, NAS-UNet, one of the first few NAS frameworks using gradient optimization to search architecture for medical image segmentation, and a conventional UNet implementation~\cite{ref_article4,ref_article7,ref_article9}. For comparison purpose, all these methods are separately trained with evaluation conducted by 5-fold cross-validation for the same metrics, while the architecture identified by MS-NAS is transferred from the searched architecture on Chaos (CT) dataset with a very small subset for tuning.
% For comparison purpose, all these methods are separately trained for each dataset to achieve its best performance and evaluated for the same metrics, while the architecture identified by MS-NAS is transferred from the searched architecture on Chaos (CT) dataset with a very small subset for tuning.

In the proposed MS-NAS implementation, the number of network layers is 10, and the number of blocks in a cell is 3, yielding a search space of $3.89\times10^9$ paths, $4.22\times10^8$ cells, and $1.64\times10^{18}$ possible network architectures. For {contracting-cell}, $maxpooling$ with $stride=2$ is used for $\frac{s}{2} \rightarrow s$ connection, while for {expanding-cell}, bilinear up-sampling is used for $2s \rightarrow s$ connection. For the partial channel connections, $k$ is set to 4. $N_l$ is varied from 3 to 5 to identify different architectures as tradeoff between accuracy and complexity, which are denoted as MS-NAS($N_l$) accordingly. 

\begin{figure}[th]
\centering\vspace{-0.5cm}
\includegraphics[width=0.85\textwidth]{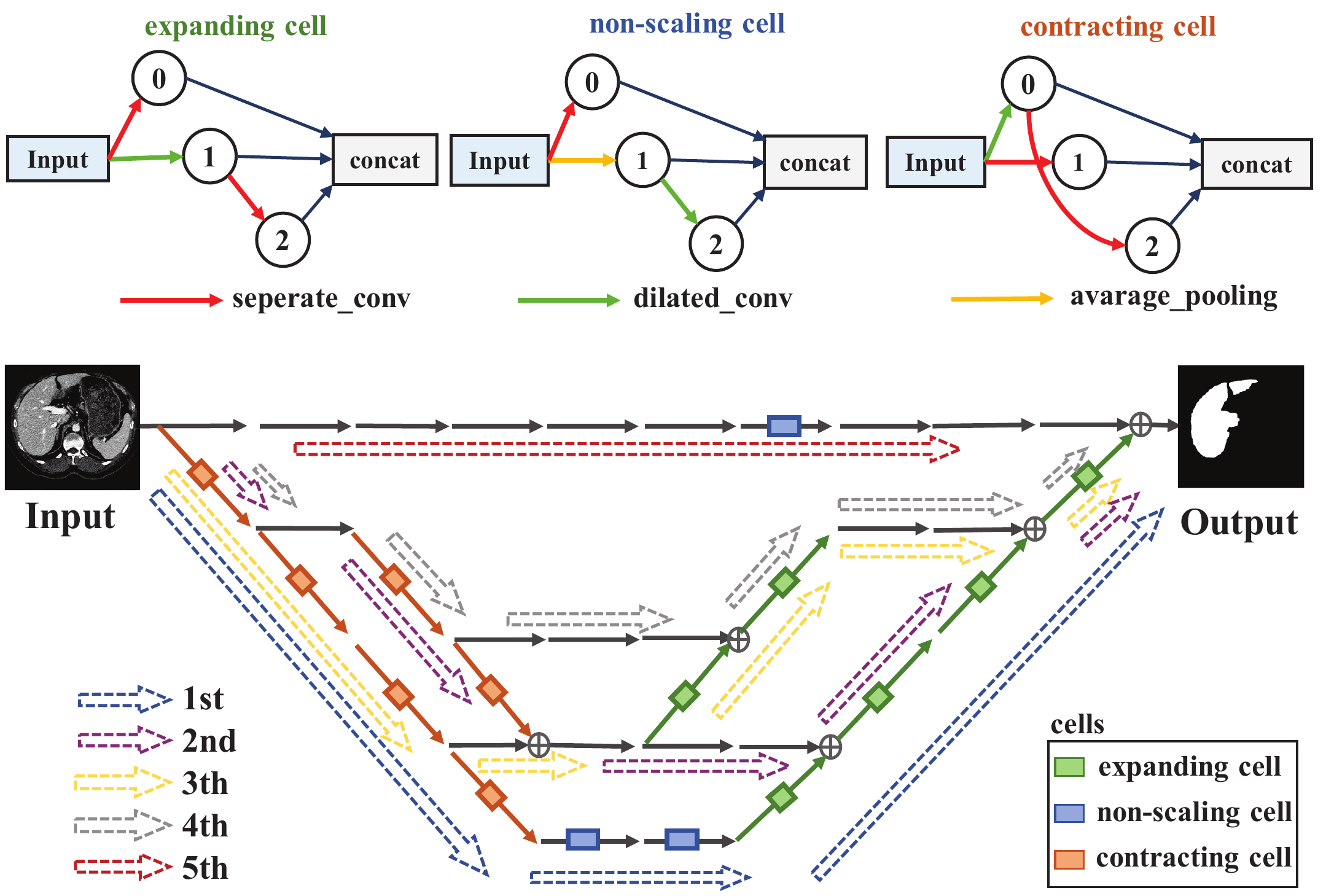}
\caption{Illustration of the optimized cell and network architecture.}\vspace{-0.9cm}
\label{fig:result}
\end{figure}

The ground-truth and the input images are resized to $256 \times 256$. A total of 40 epochs of architecture search optimization are conducted, with the first 20 epochs optimizing cell parameters and the last 20 epochs for network architecture parameters. A SGD optimizer is employed with a momentum of 0.9, learning rate from 0.025 to 0.001, and a weight decay of 0.0003. The search procedure takes about 2 days to complete on a GTX 1080Ti GPU. Fig.~\ref{fig:result} plots the optimized cell and network structure with $N_l=5$ paths, as indicated by the dashed arrows.
\subsection{Experimental Results}
\begin{figure}[ht]
\centering\vspace{-0.3cm}
\includegraphics[width=\textwidth]{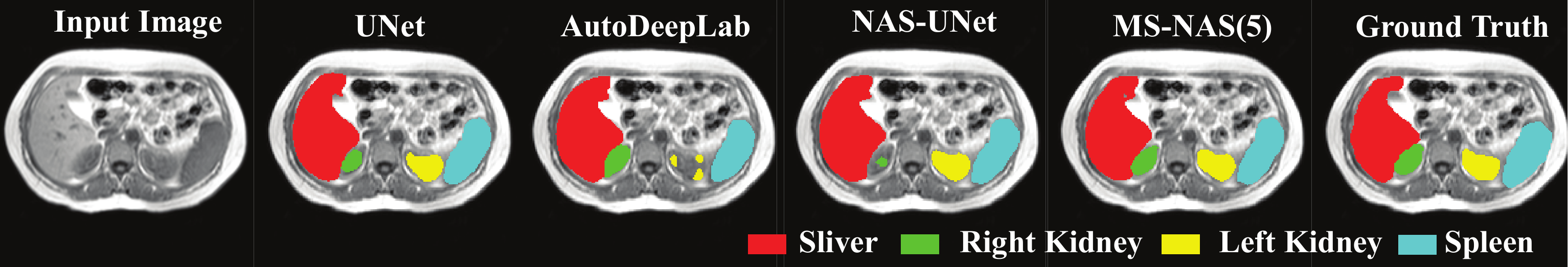}
\caption{Qualitative comparison of segmentation results for different methods.}\vspace{-0.5cm}
\label{fig:organ_show}
\end{figure}

For quantitative evaluation, Table~\ref{tab1} presents the comparison among AutoDeeplab, UNet, NAS-UNet, and the proposed MS-NAS with $N_l$ varied from 3 to 5 (denoted as MS-NAS(3) to MS-NAS(5)) on the metrics of mean Intersection over
Union (mIOU) and Dice Similarity Coefficient (DSC) for the datasets of Silver07 and Promise12. It is clear that even with $N_l$=3, MS-NAS significantly outperforms all the other frameworks. When $N_l$=5, MS-NAS can achieve 1.2-4.9\% improvement in average mIOU and 0.8-3.5\% improvement in average DSC. It is worth noting that our proposal consumes the least computational resources, with 14.0-20.3\% saving on parameter size and 18.0-24.9\% saving in computational Flops. This simply indicates MS-NAS is capable to find a more optimal architecture with reduced overhead on a larger search space. Moreover, as shown in Table~\ref{tab2}, the proposed method consistently achieves the best performance on the Chaos (MRI) dataset with 0.6-5.4\% mIOU and 0.4-3.1\% DSC improvement for MRI scans of all the organs. The hyper-parameter $N_l$ trades off between accuracy and complexity to meet different demands from different tasks while maintaining the state-of-the-art performance. Finally, an example input image and the corresponding segmentation results from the Chaos (MRI) dataset are presented in {Fig.~\ref{fig:organ_show}}, which qualitatively show consistently better segmentation performance by MS-NAS compared with other methods.

% \usepackage{multirow}
\begin{table}[t]\vspace{-0.1cm}
\centering
\tabcolsep 6pt
\scriptsize
\caption{Comparison of average mIOU, average DSC, model size, and computational flops for the datasets of Silver07 and Promise12.}\label{tab1}
\begin{tabular}{|l|c|c|c|c|c|c|}
\hline
\multicolumn{1}{|c|}{\multirow{2}{*}{Model}} & \multicolumn{2}{c|}{Sliver07} & \multicolumn{2}{c|}{Promise12} & \multirow{2}{*}{Params (M)} & \multirow{2}{*}{Flops (G)} \\ \cline{2-5}
                  &    mIOU(\%)   &      DSC(\%)  & mIOU(\%)      &    DSC(\%)    &                   &                   \\ \hline
UNet              &  95.3$\pm$0.5  &  97.6$\pm$0.6   & 65.4$\pm$0.8    & 79.1$\pm$0.9    &   13.39         &     31.01       \\ 
NAS-UNet    &  96.0$\pm$0.4   & 98.0$\pm$0.4    & 65.9$\pm$0.7    &  79.4$\pm$0.8   &   12.45         &     28.43       \\ 
AutoDeepLab &  95.2$\pm$0.6   &  97.5$\pm$0.6   &   64.2$\pm$0.9  &  78.2$\pm$0.9   &   14.45         &     33.06       \\ 
 MS-NAS(3)   &  96.7$\pm$0.4   &  98.3$\pm$0.4   &   70.1$\pm$0.6  & 82.4$\pm$0.6    &   10.52         &     21.33       \\ 
MS-NAS(4)    & 97.1$\pm$0.4    &  98.4$\pm$0.4   &   70.8$\pm$0.6  &  82.9$\pm$0.7   &   10.52         &     21.33       \\ 
 MS-NAS(5)   & \bfseries 97.2$\pm$0.3   & \bfseries 98.8$\pm$0.4   &\bfseries  70.8$\pm$0.6   & \bfseries 82.9$\pm$0.6   &   11.51         &     23.31       \\ \hline

\end{tabular}\vspace{-0.1cm}
\end{table}

\begin{table}[t]
\centering\vspace{-0.1cm}
\caption{Comparison of average mIOU and average DSC for the Chaos (MRI) dataset.}\label{tab2}
\scriptsize
\begin{tabular}{|l|c|c|c|c|c|c|c|c|}
\hline
\multicolumn{1}{|c|}{\multirow{2}{*}{Model}}  &  \multicolumn{2}{c|}{Liver} & \multicolumn{2}{c|}{Right Kidney} &   \multicolumn{2}{c|}{Left Kidney} & \multicolumn{2}{c|}{Spleen}\\
\cline{2-9}  
  &  mIOU(\%)&DSC(\%) & mIOU(\%)&DSC(\%) & mIOU(\%)&DSC(\%) & mIOU(\%)&DSC(\%) \\
\hline
UNet&         88.1$\pm$0.4&93.6$\pm$0.4 & 76.8$\pm$0.8&86.0$\pm$0.9 &    73.3$\pm$0.7&84.6$\pm$0.8 & 79.8$\pm$0.5&88.7$\pm$0.6 \\
NAS-UNet&     88.3$\pm$0.4&93.7$\pm$0.4 & 77.6$\pm$0.8&87.5$\pm$0.8 &    74.0$\pm$0.7&85.4$\pm$0.7 & 80.2$\pm$0.5&89.3$\pm$0.5  \\
AutoDeeplab&       87.9$\pm$0.6&93.5$\pm$0.6 & 75.6$\pm$0.9&85.1$\pm$1.0 &    73.1$\pm$0.8&84.2$\pm$0.9 & 78.2$\pm$0.6&87.3$\pm$0.6 \\
MS-NAS(3)&    88.7$\pm$0.4&94.0$\pm$0.5 & 78.6$\pm$0.7&88.0$\pm$0.7 &    78.2$\pm$0.7&87.7$\pm$0.8 & 81.8$\pm$0.5&89.9$\pm$0.5\\
MS-NAS(4)&    88.9$\pm$0.4&94.1$\pm$0.4 & 79.1$\pm$0.7 & 88.3$\pm$0.7 & 78.5$\pm$0.7&87.9$\pm$0.7 & \bfseries 83.0$\pm$0.5& \bfseries 90.7$\pm$0.5 \\
MS-NAS(5)& \bfseries 88.9$\pm$0.4& \bfseries 94.1$\pm$0.4& \bfseries 79.3$\pm$0.6& \bfseries 88.4$\pm$0.7 &    \bfseries 79.4$\pm$0.7& \bfseries 88.5$\pm$0.7 & 82.9$\pm$0.5&90.0$\pm$0.5 \\
\hline
\end{tabular}\vspace{-0.6cm}

\end{table}

\section{Conclusion}
\label{Conclusions}
In this paper, a multi-scale neural network architecture search framework is proposed and evaluated for medical image segmentation. In the proposed framework, multi-scale search space and multi-scale fusion of different tensor sizes are employed to achieve a larger search space and higher segmentation efficiency. To address the computational overhead caused by the larger search space, a partial channel connection scheme and a two-step decoding method are utilized to ensure high quality with reduced computational cost. Experimental results show that on various datasets with different modalities, MS-NAS can achieve consistently better performance than several state-of-the-art NAS frameworks with the least computational resource consumption.

\subsubsection{Acknowledgement.}This work was supported by National Key Research and Development Program Program of China [No.2018YFE0126300] and Key Area Research and Development Program of Guangdong Province [No.2018B030338001].

%\newpage

\end{document}